\pdfoutput=1
\documentclass[aps,prb,showpacs,superscriptaddress,numerical,reprint]{revtex4-1}
\usepackage{graphicx}
\usepackage{amsmath}
\usepackage{amssymb}
\usepackage{siunitx}
\usepackage{color}

\newcommand{\mr}{\mathrm}

\begin{document}
\title[Tunable and Switchable Coupling Between Two Superconducting Resonators]
      {Tunable and Switchable Coupling Between Two Superconducting Resonators}

\author{A.~Baust}\email{Alexander.Baust@wmi.badw-muenchen.de}
\affiliation{Walther-Mei{\ss}ner-Institut, Bayerische Akademie der Wissenschaften, D-85748 Garching, Germany}
\affiliation{Physik-Department, Technische Universit\"{a}t M\"{u}nchen, D-85748 Garching, Germany}
\affiliation{Nanosystems Initiative Munich (NIM), Schellingstra{\ss}e 4, 80799 M\"{u}nchen, Germany}
\author{E.~Hoffmann}\email{A.B.\,and E.H.\,contributed equally to this work.}
\affiliation{Walther-Mei{\ss}ner-Institut, Bayerische Akademie der Wissenschaften, D-85748 Garching, Germany}
\affiliation{Physik-Department, Technische Universit\"{a}t M\"{u}nchen, D-85748 Garching, Germany}

\author{M.~Haeberlein}
\affiliation{Walther-Mei{\ss}ner-Institut, Bayerische Akademie der Wissenschaften, D-85748 Garching, Germany}
\affiliation{Physik-Department, Technische Universit\"{a}t M\"{u}nchen, D-85748 Garching, Germany}
\author{M.~J.~Schwarz}
\affiliation{Walther-Mei{\ss}ner-Institut, Bayerische Akademie der Wissenschaften, D-85748 Garching, Germany}
\affiliation{Physik-Department, Technische Universit\"{a}t M\"{u}nchen, D-85748 Garching, Germany}
\affiliation{Nanosystems Initiative Munich (NIM), Schellingstra{\ss}e 4, 80799 M\"{u}nchen, Germany}
\author{P.~Eder}
\affiliation{Walther-Mei{\ss}ner-Institut, Bayerische Akademie der Wissenschaften, D-85748 Garching, Germany}
\affiliation{Physik-Department, Technische Universit\"{a}t M\"{u}nchen, D-85748 Garching, Germany}
\affiliation{Nanosystems Initiative Munich (NIM), Schellingstra{\ss}e 4, 80799 M\"{u}nchen, Germany}
\author{J.~Goetz}
\affiliation{Walther-Mei{\ss}ner-Institut, Bayerische Akademie der Wissenschaften, D-85748 Garching, Germany}
\affiliation{Physik-Department, Technische Universit\"{a}t M\"{u}nchen, D-85748 Garching, Germany}
\author{F.~Wulschner}
\affiliation{Walther-Mei{\ss}ner-Institut, Bayerische Akademie der Wissenschaften, D-85748 Garching, Germany}
\affiliation{Physik-Department, Technische Universit\"{a}t M\"{u}nchen, D-85748 Garching, Germany}
\author{E.~Xie}
\affiliation{Walther-Mei{\ss}ner-Institut, Bayerische Akademie der Wissenschaften, D-85748 Garching, Germany}
\affiliation{Physik-Department, Technische Universit\"{a}t M\"{u}nchen, D-85748 Garching, Germany}
\affiliation{Nanosystems Initiative Munich (NIM), Schellingstra{\ss}e 4, 80799 M\"{u}nchen, Germany}
\author{L.~Zhong}
\affiliation{Walther-Mei{\ss}ner-Institut, Bayerische Akademie der Wissenschaften, D-85748 Garching, Germany}
\affiliation{Physik-Department, Technische Universit\"{a}t M\"{u}nchen, D-85748 Garching, Germany}
\affiliation{Nanosystems Initiative Munich (NIM), Schellingstra{\ss}e 4, 80799 M\"{u}nchen, Germany}
\author{F.~Quijandr\'{i}a}
\affiliation{Instituto de Ciencia de Materiales de Arag\'{o}n and Departamento de F\'{\i}sica de la Materia Condensada, CSIC-Universidad de Zaragoza, 50009 Zaragoza, Spain}
\author{B.~Peropadre}
\affiliation{Instituto de Fisica Fundamental, IFF-CSIC, Calle Serrano 113b, Madrid E-28006, Spain}%
\affiliation{Department of Chemistry and Chemical Biology, Harvard University, Cambridge, Massachusetts 02138, United States}
\author{D.~Zueco}
\affiliation{Instituto de Ciencia de Materiales de Arag\'{o}n and Departamento de F\'{\i}sica de la Materia Condensada, CSIC-Universidad de Zaragoza, 50009 Zaragoza, Spain}
\affiliation{Fundaci\'{o}n ARAID, Paseo Mar\'{\i}a Agust\'{\i}n 36, 50004 Zaragoza, Spain}
\author{J.-J.~Garc\'{i}a Ripoll}
\affiliation{Instituto de Fisica Fundamental, IFF-CSIC, Calle Serrano 113b, Madrid E-28006, Spain}%
\author{E.~Solano}
\affiliation{Department of Physical Chemistry, University of the Basque Country UPV/EHU, Apartado 644, E-48080 Bilbao, Spain}
\affiliation{IKERBASQUE, Basque Foundation for Science, Maria Diaz de Haro 3, 48013 Bilbao, Spain}
\author{K.~Fedorov}
\affiliation{Walther-Mei{\ss}ner-Institut, Bayerische Akademie der Wissenschaften, D-85748 Garching, Germany}
\affiliation{Physik-Department, Technische Universit\"{a}t M\"{u}nchen, D-85748 Garching, Germany}
\author{E.~P.~Menzel}
\affiliation{Walther-Mei{\ss}ner-Institut, Bayerische Akademie der Wissenschaften, D-85748 Garching, Germany}
\affiliation{Physik-Department, Technische Universit\"{a}t M\"{u}nchen, D-85748 Garching, Germany}
\author{F.~Deppe}
\affiliation{Walther-Mei{\ss}ner-Institut, Bayerische Akademie der Wissenschaften, D-85748 Garching, Germany}
\affiliation{Physik-Department, Technische Universit\"{a}t M\"{u}nchen, D-85748 Garching, Germany}
\affiliation{Nanosystems Initiative Munich (NIM), Schellingstra{\ss}e 4, 80799 M\"{u}nchen, Germany}
\author{A.~Marx}
\affiliation{Walther-Mei{\ss}ner-Institut, Bayerische Akademie der Wissenschaften, D-85748 Garching, Germany}
\author{R.~Gross}\email{Rudolf.Gross@wmi.badw-muenchen.de}
\affiliation{Walther-Mei{\ss}ner-Institut, Bayerische Akademie der Wissenschaften, D-85748 Garching, Germany}
\affiliation{Physik-Department, Technische Universit\"{a}t M\"{u}nchen, D-85748 Garching, Germany}
\affiliation{Nanosystems Initiative Munich (NIM), Schellingstra{\ss}e 4, 80799 M\"{u}nchen, Germany}

\date{\today}

\begin{abstract}
We realize a device allowing for tunable and switchable coupling between two frequency-degenerate superconducting resonators mediated by an artificial atom. For the latter, we utilize a persistent current flux qubit. We characterize the tunable and switchable coupling in frequency and time domain and find that the coupling between the relevant modes can be varied in a controlled way. Specifically, the coupling can be tuned by adjusting the flux through the qubit loop or by controlling the qubit population via a microwave drive. Our measurements allow us to find parameter regimes for optimal coupler performance and quantify the tunability range.
\end{abstract}


\maketitle
\section{Introduction}
Circuit quantum electrodynamics (QED) \cite{Wallraff:2004a} has become a well-established platform for the investigation of light-matter interaction,\cite{niemczyk_circuit_2010} quantum information processing and, recently, quantum simulation.\cite{Houck2012, Raftery2013, Chen2014weak} One of the most important advantages in using superconducting circuits for these purposes is the large coupling strength between the main building blocks, namely superconducting quantum bits and microwave resonators. Noticeably, the coupling strength remains considerable even for second-order mechanisms. However, to realize quantum gates and quantum information and simulation protocols, the coupling between the individual circuit elements needs to be tunable \emph{in situ}. This can be realized in at least two ways. One way is to decouple two circuits by detuning them in frequency, for example by using the frequency tunability of superconducting qubits. With this technique, systems with up to five qubits and up to five microwave resonators were studied, \cite{Lucero2012, Barends2014, Chen2014weak} entangled quantum states were created \cite{Eichler2012, Saira2014, DiCarlo2010} and quantum teleportation \cite{Steffen2013} and quantum computing protocols were demonstrated.\cite{Dewes2012, dewes_characterization_2012, Reed2012} Alternatively, the coupling between two circuit QED building blocks can be mediated by additional coupling circuits. Examples for coupling circuits include single Josephson junctions,\cite{Chen:2014, Bialczak:2011, Flurin2014} SQUIDs \cite{vanderPloeg2007, Hime2006, Yin:2013, Peropadre2013, Pierre2014, Allman2014} or qubits \cite{Niskanen2006, Niskanen2007, Hoi2013} which were used to realize tunable coupling between qubits, resonators and transmission lines. Furthermore, new types of qubits were introduced featuring intrinsic tunability of the coupling to microwave resonators.\cite{Hoffman2011, Srinivasan2011, Gambetta2011, Makhlin1999} In this work, we report on tunable and switchable coupling between two frequency-degenerate superconducting transmission line resonators mediated in a second-order process by a superconducting flux qubit.\cite{Mariantoni2008, reuther_two-resonator_2010} Our setup is in a way dual to the usage of a resonator as quantum bus between two qubits.\cite{Sillanpaa2007, majer_coupling_2007} One particular property of our scheme is that the coupling between the two resonators can either be tuned via the magnetic flux applied to the qubit loop or switched by varying the qubit population via a microwave drive. We perform time domain measurements to find the parameter regimes for optimal sample performance. We point out that tunable coupling between frequency-degenerate resonators is of particular importance in the light of recent proposals on quantum simulations of many-body physics.\cite{Schmidt2010, Angelakis2007, Raftery2013, Hartmann2010, Houck2012, Leib2012} All these proposals and experiments would obviously profit from a well-controlled tunable resonator-resonator coupling such as the one presented in this work.
\\

\section{Sample and Measurement Setup}
\begin{figure}[t]
    \includegraphics{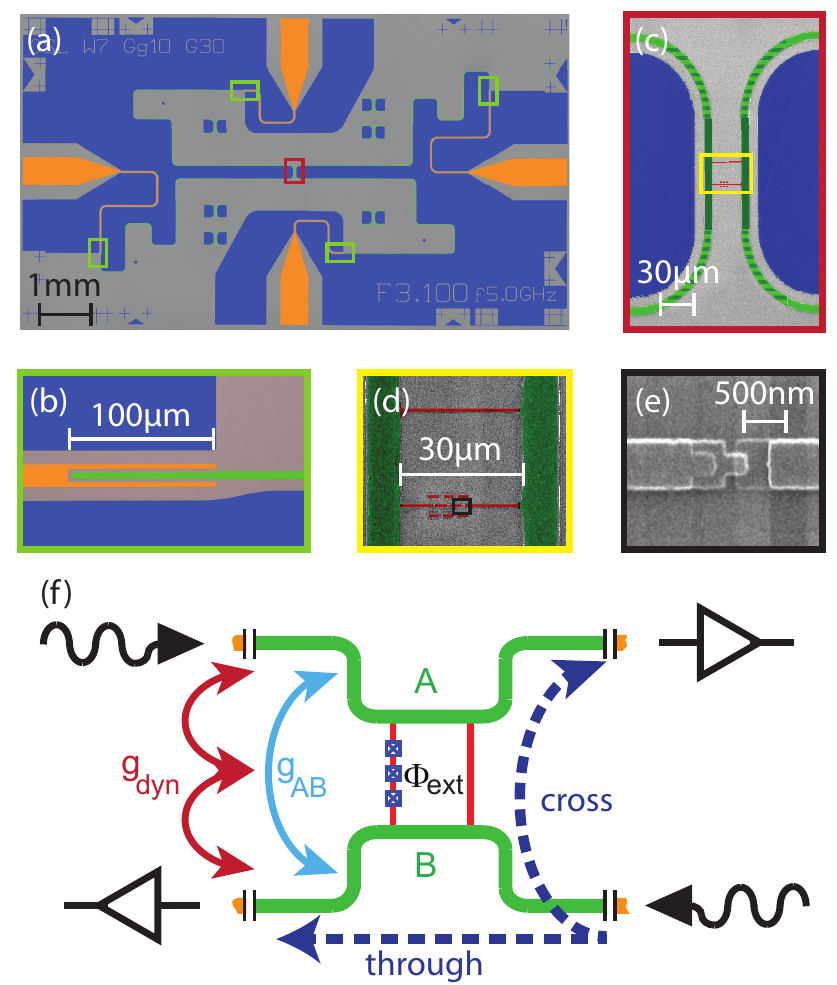}
   \caption{(color online) (a) False-color image of the sample. Nb ground planes are shown in blue and feed lines in orange. The resonator signal lines reside along the ground plane edges. (b) Coupling capacitor defining the resonators. (c) Resonator coupling area with signal lines (green) and flux qubit (red). Light/dark green stripes highlight Nb-Al overlap areas. (d) Flux qubit galvanically coupled to both resonators. (e) Al/$\rm{AlO_x}$/Al Josephson junction fabricated using shadow evaporation. (f) Working principle of the coupler and measurement setup.}
  \label{Hoffmann_Fig1}
\end{figure}

Our sample comprises two coplanar stripline resonators, A and B, with fundamental mode frequencies $\omega_{\rm{R}}/2\pi\,{=}\,4.896\,\mr{GHz}$ and a superconducting flux qubit as artificial atom as shown in Figs.\,\ref{Hoffmann_Fig1}(a)-(c). The resonators are fabricated in Nb technology on a thermally oxidized Si substrate. The linewidths of the fundamental modes of both resonators for the qubit being far detuned were determined as $\gamma_\mr{A}/2\pi\,{=}\,2.3\,\mr{MHz}$ and $\gamma_\mr{B}/2\pi\,{=}\,0.5\,\mr{MHz}$. The detuning between the two resonators is found to be small and is therefore neglected. An artificial atom is coupled galvanically to the signal lines of both resonators at the position of the current antinodes of their fundamental modes, cf.~Figs.\,\ref{Hoffmann_Fig1}(c)-(e). In our case, this artificial atom is a flux qubit consisting of a superconducting Al loop with three Josephson junctions where one of the junctions is smaller by a factor $\alpha\,{\simeq}\,0.7$. For the qubit we determine an energy gap $\Delta/h\,{=}\,3.55\,\mr{GHz}$ and a persistent current $I_{\rm{p}}\,{=}\,458\,\mr{nA}$. The coupling between the qubit and each resonator is $g/2\pi\,{=}\,96.7\,\mr{MHz}$. The qubit parameters determined by two-tone spectroscopy can be quantitatively described by taking into account the galvanic coupling of the qubit to the resonators in our setup, see appendix.
\\

The effective Hamiltonian for the qubit coupled to the fundamental modes of the two resonators is \cite{Mariantoni2008, reuther_two-resonator_2010}
\begin{equation}\label{eq:QS_eff}
    \begin{split}
        \hat{H}_\mr{eff} =  \hbar\frac{\omega_\mr{Q}}{2}\hat{\sigma}_z &+\hbar\left(\omega_\mr{R}+ g_\mr{dyn}\hat{\sigma}_z\right)\left(\hat{a}^\dag\hat{a}+\hat{b}^\dag\hat{b}\right)\\
        & +\hbar\left(g_\mr{AB}+g_\mr{dyn}\hat{\sigma}_z\right)\left(\hat{a}^\dag\hat{b}+\hat{a}\hat{b}^\dag\right).
    \end{split}
\end{equation}
Here, $\omega_{\rm{Q}}\,{=}\,\sqrt{\epsilon^2+\Delta^2}/\hbar$ is the qubit transition frequency with the energy bias $\epsilon(\Phi_{\rm{ext}})\,{=}\,2I_{\rm{p}}(\Phi_{\rm{ext}}-\Phi_0/2)$, $\Phi_0$ is the flux quantum, and $\Phi_{\rm{ext}}$ is the external magnetic flux threading the qubit loop. At the flux degeneracy point $\delta\Phi_\mathrm{ext}\,{=}\,\Phi_{\rm{ext}}\,{-}\,\Phi_0/2\,{=}\,0$, one finds $\hbar\omega_{\rm{Q}}(\Phi_0/2)\,{=}\,\Delta$. Furthermore, we denote the annihilation (creation) operators for the two resonators A and B as $\hat{a}$ and $\hat{b}$ ($\hat{a}^\dagger$ and $\hat{b}^\dagger$), respectively. The coupling between the two resonators is mediated by two mechanisms. In addition to the geometric coupling $g_{\rm{AB}}/2\pi\,{=}\,8.4\,\rm{MHz}$  between the two resonators there is the flux-dependent second-order \textit{dynamic coupling} $g_\mr{dyn}\langle\hat{\sigma}_z\rangle\,{\equiv}\, (g\sin\theta)^2[(\omega_\mr{Q}\,{-}\,\omega_\mr{R})^{-1}\,{+}\,(\omega_\mr{Q}\,{+}\,\omega_\mr{R})^{-1}]\langle\hat{\sigma}_z\rangle$.
As a consequence, the total resonator-resonator coupling
$$
g_\mr{res} \equiv g_\mr{AB} + g_\mr{dyn}\langle\hat{\sigma}_z\rangle
$$
can be tuned via $\Phi_{\rm{ext}}$ since both the mixing angle $\tan\theta\,{=}\,\Delta/\epsilon$ and the qubit transition frequency $\omega_{\rm{Q}}$ are flux dependent.
We gain further insight by considering the normal modes of the coupled resonators  $\hat{c}_\pm\,{=}\,\frac{1}{\sqrt{2}}\left(\hat{a}\pm\hat{b}\right)$ and $\hat{c}^\dag_\pm\,{=}\,\frac{1}{\sqrt{2}}\left(\hat{a}^\dag\pm\hat{b}^\dag\right)$,
which allow us to rewrite the Hamiltonian of Eq.\,(\ref{eq:QS_eff}) to
\begin{align}\label{eq:rotHam}
        \hat{H}_\mr{eff} = & \,\hbar\frac{\omega_\mr{Q}}{2}\hat{\sigma}_z +
        \hbar\omega_\mr{R}(\hat{c}_+^\dag\hat{c}_++ \hat{c}_-^\dag\hat{c}_-)\nonumber\\
        &+ \hbar g_\mr{AB}(\hat{c}_+^\dag\hat{c}_+- \hat{c}_-^\dag\hat{c}_-) +
        2\hbar g_\mr{dyn}\;\hat{\sigma}_\mr{z} \hat{c}_+^\dag\hat{c}_+
\end{align}
The modes $\hat{c}_{-}$ and $\hat{c}_{+}$ correspond to in-phase and out-of-phase oscillating currents in the two resonators, respectively. As only the out-of-phase oscillating mode generates a magnetic field at the position of the qubit, only this mode couples to the qubit.
\\

For our measurements, we mount the sample inside a gold-plated copper box to the base temperature stage of a dilution refrigerator. The sample temperature is stabilized at $45\,\rm{mK}$. As shown in Fig.\,\ref{Hoffmann_Fig1}(f), one port of each resonator is connected to a highly attenuated input line while the respective second port is connected to an output line featuring cryogenic and room temperature amplifiers. In this way, we can measure the transmission through the individual resonators (referred to as a `through'-measurement) but also the transmission from the input of one resonator to the output of the second resonator (`cross'-measurement). A superconducting solenoid is mounted on top of the sample package in order to apply magnetic flux to the qubit loop.
\begin{figure*}[t]
    \includegraphics{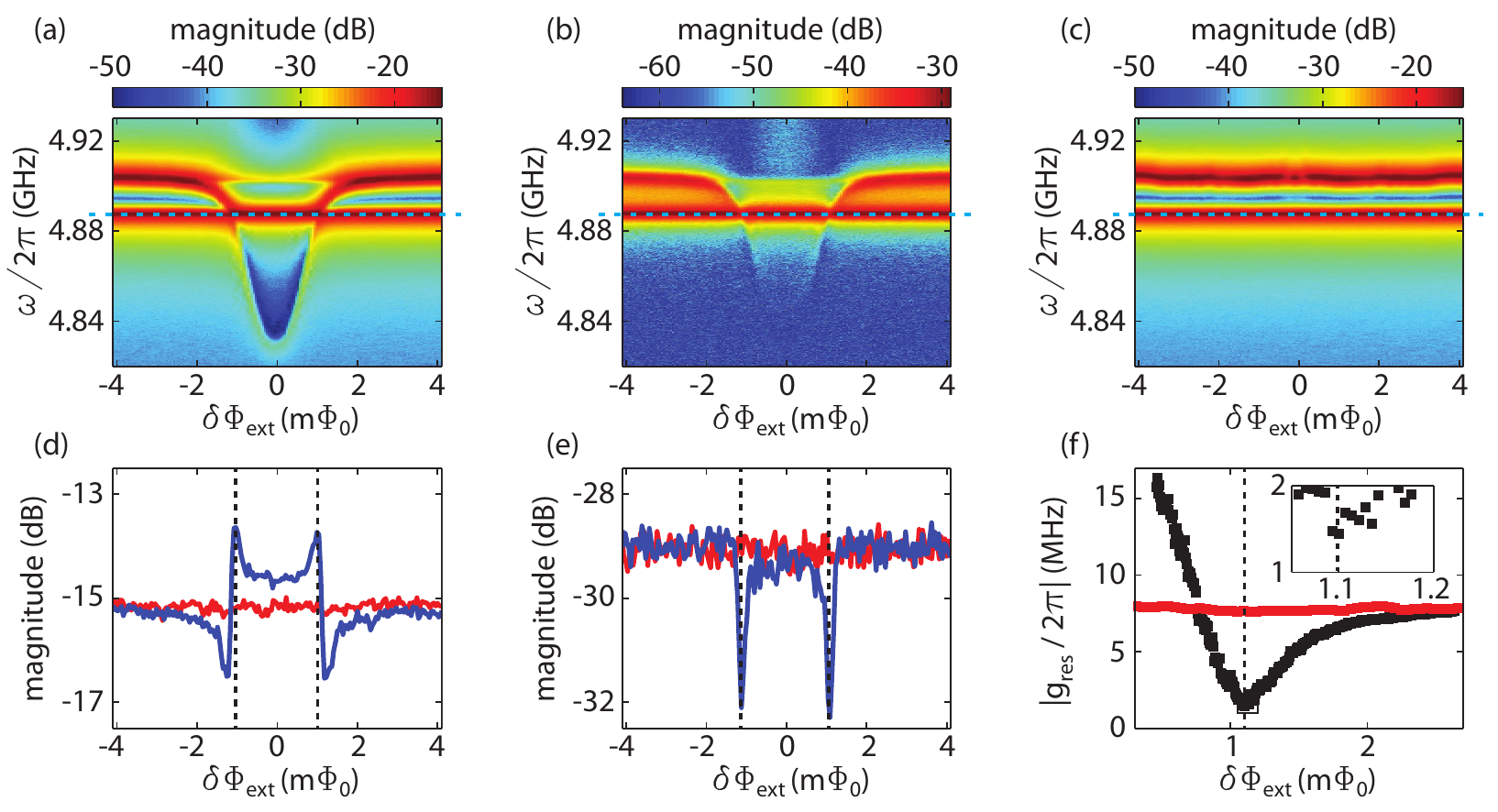}
   \caption{(color online) (a) Transmission through one resonator depending on the applied magnetic flux with the qubit in ground state. (b) `Cross' measurement, qubit in ground state. (c) `Through' measurement, qubit driven with strong excitation signal. (d) `Through' measurement, transmission at the frequency of the lower mode at $4.888\,\rm{GHz}$ [dashed lines in (a) and (c)] with qubit in ground state (blue line) and saturated qubit (red line). Dashed black lines: Switch setting conditions. (e) Same as (d) for the `cross' measurement. (f) Magnitude of the total coupling between the resonators extracted from a `through' measurement near the switch setting condition. Black:\,qubit in ground state. Red:\,saturated qubit. Inset:\,Measurement with increased flux resolution around switch setting condition.}
  \label{Hoffmann_Fig2}
\end{figure*}

\section{Results}
In this section, we investigate two ways of controlling the coupling between the two resonators: first, via the external magnetic field and, second, via the qubit population.
\subsection{Tuning the coupling via the external field}
To determine the relevant sample parameters and to characterize the coupler properties, we first measure the transmission through the resonators with a vector network analyzer as a function of the applied magnetic flux $\Phi_{\rm{ext}}$. Fig.\,\ref{Hoffmann_Fig2}(a) shows the results of a `through'-measurement whereas Fig.\,\ref{Hoffmann_Fig2}(b) represents a `cross'-measurement. For both measurements, the input signal is applied to the same port and the qubit remains in the ground state. The input power is chosen such that the population of both resonators is approximately one photon on average. We observe two modes as expected for coupled resonators, where the splitting far away from the qubit degeneracy point is $2g_{\rm{AB}}$. If the flux is tuned towards the degeneracy point, the frequency of the lower mode stays constant while the frequency of the upper mode is shifted to lower frequencies as expected from Eq.\,(\ref{eq:rotHam}).

In this way, the flux can be tuned such that the frequency of the upper mode matches the frequency of the lower mode. We refer to these points as the \textit{switch setting conditions} where the geometric coupling is fully compensated by the dynamical coupling. Consequently, the two resonators are expected to be decoupled from each other if the switch setting condition is fulfilled. In order to find the minimum value of the coupling for our device, we fit the mode spectrum shown in Fig.\,\ref{Hoffmann_Fig2}(a) using input-output theory~\cite{walls2008quantum, Sames2014} and analyze the coupling depending on the magnetic flux. Results are shown in Fig.\,\ref{Hoffmann_Fig2}(f). At the switch setting condition, the coupling is reduced to $|g_{\mr{res,min}}/2\pi|\,{\lesssim}\,1.5\,\mr{MHz}$. Here, our analysis is limited by the decay rates of the resonators. Compared to the coupling far off the degeneracy point, the coupling at the switch setting condition is reduced by a factor of at least $5.5$.

\subsection{Tuning the coupling via the qubit population}
So far, we have investigated how to tune the coupling via the magnetic flux applied to the qubit loop. Next, we show that the coupling is also controlled by the qubit population as expected from Eq.\,(\ref{eq:rotHam}). To this end, we record the resonator transmission while driving the qubit with a strong excitation signal applied through the input port of the other resonator. This results in equal probabilities to find the qubit in the ground and excited state, yielding $\langle\hat{\sigma}_{\rm{z}}\rangle\,{=}\,\rm{Tr}\left[\rho_{\rm{M}}\hat{\sigma}_{\rm{z}}\right]\,{=}\,0$ where $\rho_{\rm{M}}\,{=}\,\frac{1}{2}\left(\left|\rm{g}\right\rangle\left\langle\rm{g}\right|+ \left|\rm{e}\right\rangle\left\langle\rm{e}\right|\right)$. As expected from Eq.\,(\ref{eq:QS_eff}) we observe that the coupling between the two resonators is then given by $g_{\rm{AB}}$ independently of the applied flux, see Fig.\,\ref{Hoffmann_Fig2}(c) and Fig.\,\ref{Hoffmann_Fig2}(f). To analyze the interplay of flux- and qubit state dependence in more detail, we show the transmission at the frequency of the lower mode at $\omega/2\pi\,{=}\,4.888\,\mr{GHz}$ in Fig.\,\ref{Hoffmann_Fig2}(d) and Fig.\,\ref{Hoffmann_Fig2}(e). For the qubit in the ground state, we observe increased transmission for the `through'-measurement at the switch setting conditions compared to flux values not matching a switch setting condition or compared to the qubit being driven. This is in agreement with our expectation that, when turning off the coupling, the signal incident on one resonator cannot cross over to the other one. Consistently, we observe reduced transmission at the switch setting condition in the `cross' measurement shown in Fig.\,\ref{Hoffmann_Fig2}(e). Two dips are visible in the through transmission [Fig.\,\ref{Hoffmann_Fig2}(d)], when the qubit is in the ground state. They originate from the differences in the linewidths and also from a possible small detuning between the two resonators. The resonant structure close to the frequency of the out-of-phase mode [cf.~Fig.\,\ref{Hoffmann_Fig2}(a) and Fig.\,\ref{Hoffmann_Fig2}(b)], is suppressed by approx.~$15\,\mr{dB}$ and not relevant for the discussion presented here.
\begin{figure}[t]
    \includegraphics{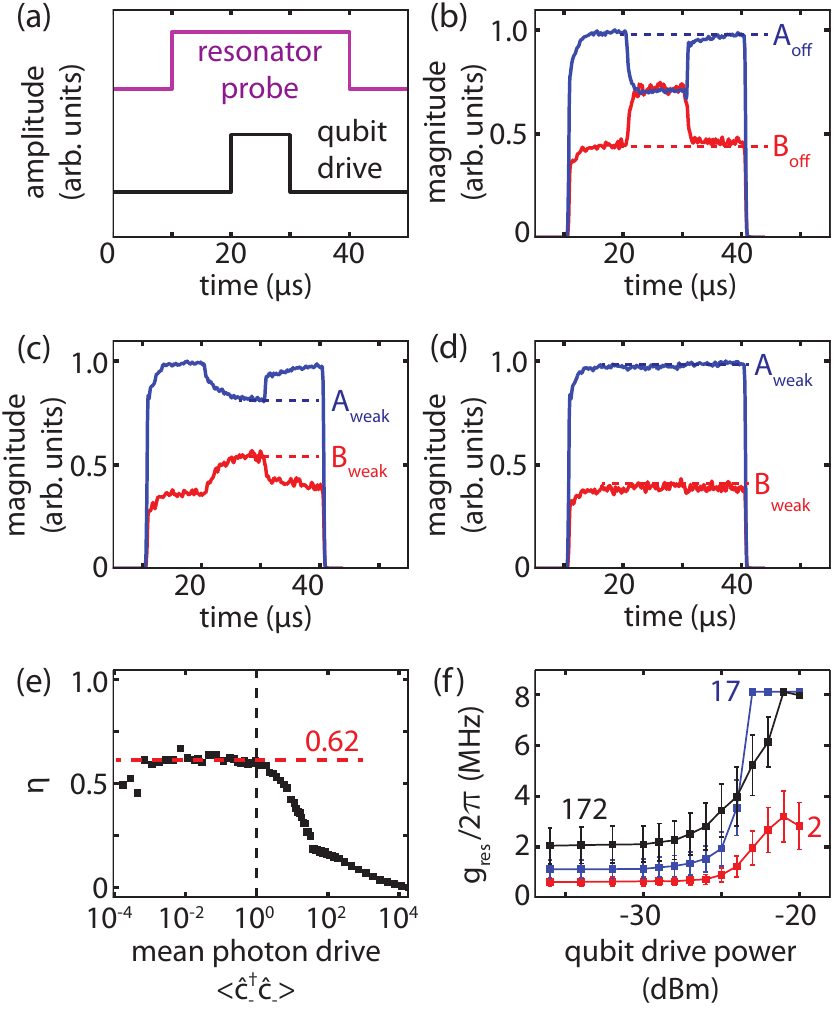}
   \caption{(color online) (a) Pulse pattern for the time-domain probe of the coupler. (b) Typical measured time traces of the output signals of the two resonators. The qubit drive pulse is strong enough to  saturate the qubit. Blue:\,`through' transmission measurement. red:\,`cross' transmission measurement. The power levels are referred to the insides of the resonators, i.e., they are scaled such that they are equal when the coupling is `on'. This assumption is justified because $g_{\mr{AB}}\,{\gg}\,\gamma_\mr{A},\gamma_\mr{B}$. (c) Same as (b) for intermediate qubit drive pulse power. (d) Same as (b) for small drive pulse power. (e) Switching efficiency $\eta$ as a function of the mean resonator drive. (f) Total resonator-resonator coupling as a function of the qubit drive power (referenced to signal generator output) measured for three different resonator populations. The points of the red curve at $-20\,\mr{dBm}$, $-24\,\mr{dBm}$, $-30\,\mr{dBm}$ are derived from the data of (b), (c), (d), respectively.}
  \label{Hoffmann_Fig3}
\end{figure}

Next, we conduct time domain experiments making the switchable coupling directly observable. To this end, we set the flux bias corresponding to the switch setting condition and apply a microwave probe pulse (length $\tau_\mathrm{res}\,{=}\,\SI{30}{\micro\second}$) to one of the resonators at the frequency $\omega_\mathrm{res}/2\pi\,{=}\,4.888\,\mr{GHz}$ of the lower ($\hat{c}_-$) mode. In addition, a $\SI{10}{\micro\second}$ long microwave driving pulse switches the coupling between the resonators on for a period of $\SI{10}{\micro\second}$ as shown in Fig.\,\ref{Hoffmann_Fig3}(a). The output signals of both resonators are detected in a time-resolved way using FPGA-enhanced A/D-converters. Typical pairs of time traces are shown in Fig.\,\ref{Hoffmann_Fig3}(b-d). After switching on the qubit drive, the output signal level of the resonator where the probe pulse is applied decreases, whereas it increases for the other resonator. This result represents a direct experimental evidence for the expected switching behaviour because it implies that the transfer of energy from one into another resonator can be controlled via the qubit.
However, for an ideal coupler, one would expect that at the switch setting condition the output signal level for the `cross'-measurement is zero when the qubit is in the ground state, even if the probe pulse is on. Nevertheless, in our case a finite output power can be observed. We attribute this to the complex mode structure of our particular device, cf.\,Fig.\,\ref{Hoffmann_Fig2}(a),  Fig.\,\ref{Hoffmann_Fig2}(b) and the appendix.

To quantify the coupler performance, we define the switching efficiency $\eta\,{\equiv}\,1\,{-}\,n_\mathrm{B}^\mathrm{off}/n_\mathrm{A}^\mathrm{off}\,{=}\,1\,{-}\,B_\mr{off}/A_\mr{off}$. Here and in the following, $n_\mathrm{B}^\mathrm{on/off}$ and $n_\mathrm{A}^\mathrm{on/off}$ denote the resonator populations when the coupling is switched on/off. Following input-output theory~\cite{walls2008quantum}, the ratio $n_\mathrm{B}^\mathrm{off}/n_\mathrm{A}^\mathrm{off}$ is equal to that of the quantities $B_\mr{off}$ and $A_\mr{off}$ indicated in Fig.\,\ref{Hoffmann_Fig3}(b).
The switching efficiency $\eta$ is most intuitively understood by looking at its limiting cases. For a perfect coupler ($\eta\,{=}\,1$), we find $n_\mathrm{A}^\mathrm{on}\,{=}\,n_\mathrm{B}^\mathrm{on}$ when the coupling is switched on and $n_\mathrm{A}^\mathrm{off}\,{=}\,n$, $n_\mathrm{B}^\mathrm{off}\,{=}\,0$ when the coupling is switched off. Conversely, when the coupler is not tunable at all ($\eta\,{=}\,0$), $n_\mathrm{A}^\mathrm{on/off}\,{=}\,n_\mathrm{B}^\mathrm{on/off}$ regardless of the coupler state. For intermediate values of $\eta$, a fraction of $(1\,{-}\,\eta)/(2\,{-}\,\eta)$ photons leaks into resonator B despite the coupler being in the 'off' state.

Next, we analyze $\eta$ as a function of mean number of photons (calibrated via dispersive shift of the qubit; data not shown) in the $\hat{c}_-$-mode. The results are shown in Fig.\,\ref{Hoffmann_Fig3}(e). For low photon numbers we find a switching efficiency of $\eta\approx 0.62$. Above approximately $1$ photon, $\eta$ starts to decrease and vanishes for photon numbers exceeding $10^4$. This behaviour is in agreement with the disappearance of the Jaynes-Cummings-doublet due to a quantum-to-classical transition observed in a transmon-resonator system.\cite{Fink2010}

Finally, we demonstrate that the resonator-resonator coupling strength can also be controlled via the qubit drive power, cf.\,Fig.\,\ref{Hoffmann_Fig3}(f). This scenario is of particular importance for the simulation of, e.g., the Bose-Hubbard-Hamiltonian where it is favorable to be able to vary the coupling between adjacent resonators by an easily controllable external parameter such as the qubit drive power. For a given qubit drive pulse power and mean resonator photon number, we find the corresponding resonator-resonator coupling by comparing the output powers of both resonators found in our measurements with the output fields expected from input-output-theory. For low resonator probe photon numbers and weak qubit drive, the residual coupling between the resonators is determined as $\SI{0.62 \pm 0.16 }{\mega\hertz}$, representing a reduction of the coupling strength by one order of magnitude as compared to the geometric coupling $g_\mr{AB}$. The error bars in Fig.\,\ref{Hoffmann_Fig3}(f) account for small detunings between the resonator probe signal frequency and the frequency of the lower switch mode $\hat{c}_{-}$. For strong qubit driving, the resonator-resonator coupling increases and converges towards the geometric coupling $g_\mr{AB}$. We note that for high qubit drive powers, the calculated coupling rates are very sensitive to small uncertainties in the quantities $A_\mr{weak}$ and $B_\mr{weak}$ [cf.\,Fig.\,\ref{Hoffmann_Fig3}(c) and Fig.\,\ref{Hoffmann_Fig3}(d)] since the mean resonator population becomes independent of the coupling rate $g_\mr{res}$ as soon as $g_\mr{res}\gg\gamma_\mr{A},\gamma_\mr{B}$.

\section{Conclusions}
In conclusion, we present a device allowing to tune the coupling between two coplanar stripline resonators via a flux qubit coupled to both of them. We characterize the individual constituents and the switching behaviour by means of spectroscopy and perform a quantitative analysis of the coupler performance using a time domain experiment. From the latter experiments, we find a coupling range of $\SI{0.62}{\mega\hertz}\,{\le}\,g_\mr{res}/2\pi\,{\le}\,\SI{8.4}{\mega\hertz}$. This corresponds to a maximum switching efficiency of $62\%$. Improved designs are promising candidates for applications in future quantum information processing setups where our coupler can be used for the controlled transfer of excitations between a fast bus resonator, to which additional qubits can be coupled, and a long-lived storage resonator serving as quantum memory. Furthermore, even with its current performance, our coupler may become a key element in quantum simulation architectures such as chains or networks of superconducting nonlinear resonators for the simulation of the Bose-Hubbard-Hamiltonian.\cite{Schmidt2010, Angelakis2007, Raftery2013, Hartmann2010, Houck2012, Leib2012}

\section*{Acknowledgements}
This work is supported by the German Research Foundation through SFB 631, EU projects CCQED, PROMISCE and SCALEQIT, Spanish Mineco projects FIS2012-33022 and FIS2012-36673-C03-02, UPV/EHU UFI 11/55 and Basque Government IT472-10. B.~P.~acknowledges support from the STC Center for Integrated Quantum Materials, NSF Grant No.~DMR-1231319.

\section*{Appendix: Fit of the spectroscopy data}
\begin{figure}[htb]
    \includegraphics{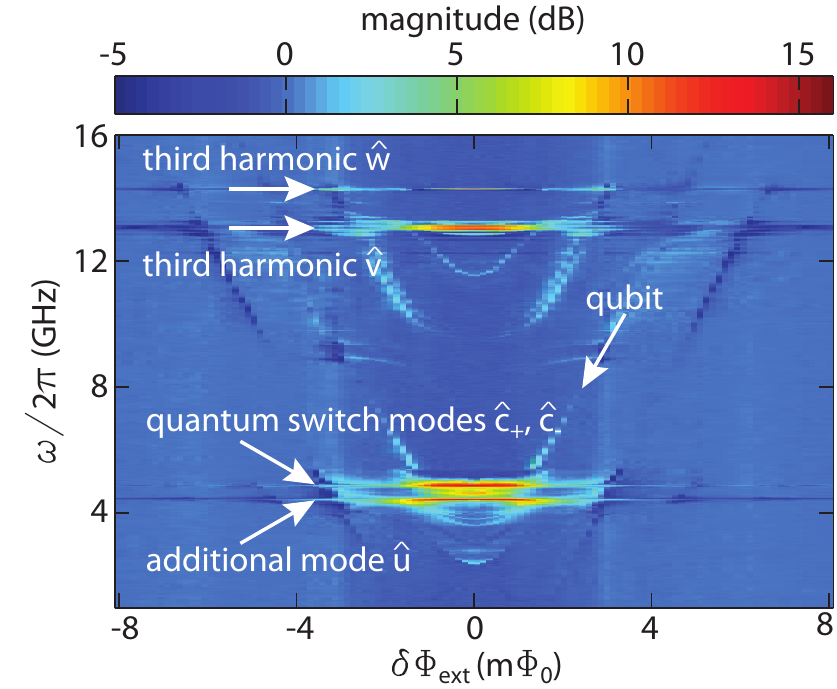}
   \caption{(color online) Two-tone spectroscopy of the sample. The switch modes $\hat{c}_+$, $\hat{c}_-$ and an additional mode $\hat{u}$ at $4.5\,\mr{GHz}$ can be identified. At $\omega_4/2\pi\,{=}\,13.1\,\mr{GHz}$ and $\omega_5/2\pi\,{=}\,14.3\,\mr{GHz}$ the third harmonics $\hat{v}$ and $\hat{w}$ of these modes are observed.}
  \label{Baust_Supplementary_Twotone}
\end{figure}

To determine the switch parameters, we fit the effective switch Hamiltonian to our spectroscopy data. However, as shown in Fig.\,\ref{Baust_Supplementary_Twotone}, there exists an additional mode $\hat{u}$ at $\omega_3/2\pi\,{=}\,4.5\,\mr{GHz}$ which couples to the qubit and therefore needs to be taken into account. To increase precision, we also include the third harmonic of this mode (denoted by $\hat{v}$, located at $\omega_4/2\pi\,{=}\,13.1\,\mr{GHz}$) and the third harmonic of the $\hat{c}_+$-mode, denoted by $\hat{w}$,  at $\omega_5/2\pi\,{=}\,14.3\,\mr{GHz}$. The third harmonic mode frequencies were found using two-tone spectroscopy,\cite{niemczyk_circuit_2010} see Fig.\,\ref{Baust_Supplementary_Twotone}. We note that we do not consider the second harmonics since they exhibit current nodes at the qubit position and therefore do not couple to the qubit.


The Hamiltonian taking the switch modes and all additional modes into account then reads
\begin{align}\label{eq:rotHamextended}
        \hat{H}_\mr{eff} &= \frac{\varepsilon}{2}\hat{\sigma}_z + \frac{\Delta}{2}\hat{\sigma}_x \nonumber\\
        &+ \hbar\omega_\mr{+}\;\hat{c}_+^\dag\hat{c}_+ + \hbar\omega_\mr{-}\;\hat{c}_-^\dag\hat{c}_- \nonumber\\
        &+ \hbar g\sqrt{2}\;\hat{\sigma}_z(\hat{c}_+^\dag\,+\,\hat{c}_+) \nonumber\\
        &+ \hbar\omega_3\;\hat{u}^\dag\hat{u} + \hbar g_3\; \hat{\sigma}_z (\hat{u}^\dag + \hat{u})\nonumber\\
        &+ \hbar\omega_4\;\hat{v}^\dag\hat{v} + \hbar g_4\; \hat{\sigma}_z (\hat{v}^\dag + \hat{v})\nonumber\\
        &+ \hbar\omega_5\;\hat{w}^\dag\hat{w} + \hbar g_5\; \hat{\sigma}_z (\hat{w}^\dag + \hat{w}).
\end{align}
From the fit, we get the following set of parameters:
\begin{figure}[htb]
    \includegraphics{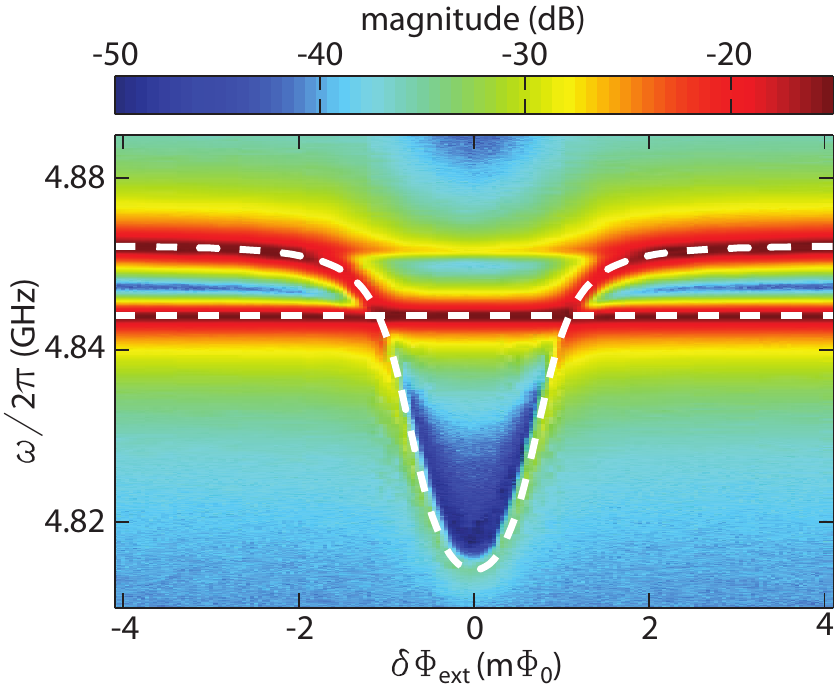}
   \caption{(color online) Transmission through one resonator as shown in the main text. Dashed lines: Fit of the complete Hamiltonian (\ref{eq:rotHamextended}) to the data.}
  \label{Baust_Supplementary_QSwithfit}
\end{figure}
\begin{align*}\label{eq:params}
        \Delta/h &= 3.55\,\mr{GHz}\nonumber\\
        I_\mr{p} &= 458\,\mr{nA}\nonumber\\
        \omega_+/2\pi &= 4.9044\,\mr{GHz}\nonumber\\
        \omega_-/2\pi &= 4.888\,\mr{GHz}\nonumber\\
        \omega_3/2\pi &= 4.5\,\mr{GHz}\nonumber\\
        \omega_4/2\pi &= 13.1\,\mr{GHz}\nonumber\\
        \omega_5/2\pi &= 14.3\,\mr{GHz}\nonumber\\
        g/2\pi &= 96.7\,\mr{MHz}\nonumber\\
        g_3/2\pi &= 775\,\mr{MHz}\nonumber\\
        g_4/2\pi &=g_3/2\pi\cdot\sqrt{\frac{\omega_4}{\omega_3}}= 1323\,\mr{MHz}\nonumber\\
        g_5/2\pi &=g/2\pi\cdot\sqrt{2}\cdot\sqrt{\frac{\omega_5}{\omega_+}}= 233\,\mr{MHz}\nonumber\\
        g_\mr{AB} &= 8.4\,\mr{MHz}\nonumber\\
\end{align*}

 Instead of using $g_4$ and $g_5$ as independent fit parameters, we calculate the coupling of the third harmonics using the ratio of the resonant frequencies of the third and fundamental modes. For $g_5$, the factor of $\sqrt{2}$ arises from the fact that the coupling strength of the $c_+$-mode is given by $\sqrt{2}\cdot g$. Fig.\,\ref{Baust_Supplementary_QSwithfit} shows the two switch modes together with the fit. As can be seen, experimental data and theory correspond very well.

%


\end{document}